\documentclass[preprintnumbers,amsmath,amssymb,showpacs]{revtex4}

\usepackage{graphicx}
\usepackage{dcolumn}
\usepackage{bm}
\usepackage{booktabs}
\usepackage{CJK}

\begin{document}

\title{Exploration of multiphoton entangled states by using weak nonlinearities}

\author{Yingqiu He $^1$, Dong Ding $^{1,2}$, Fengli Yan $^{1}$}
\email{flyan@hebtu.edu.cn}
\author{Ting Gao $^3$}
\email{gaoting@hebtu.edu.cn}
\affiliation {$^1$ College of Physics Science and Information Engineering, Hebei Normal University, Shijiazhuang 050024, China \\
$^2$Department of Basic Curriculum, North China Institute of Science and Technology, Beijing 101601, China \\
$^3$College of Mathematics and Information Science, Hebei Normal University, Shijiazhuang 050024, China}
\date{\today}

\begin{abstract}

We propose a fruitful scheme for exploring multiphoton entangled states based on linear optics and weak nonlinearities.
Compared with the previous schemes the present method is more feasible because there are only small phase shifts instead of a series of related functions of photon numbers in the process of interaction with Kerr nonlinearities.
In the absence of decoherence we analyze the error probabilities induced by homodyne measurement and show that the maximal error probability can be made small enough even when the number of photons  is large.
This implies that the present scheme is quite tractable and it is possible to produce entangled states involving a large number of photons.

\end{abstract}

\maketitle

\section{Introduction}

Undoubtedly, entanglement \cite{HHHH2009, GT2009, YGC2011, GYE2014} is one of the most crucial elements in quantum information processing. In recent years, quantum entanglement has been extensively investigated in various candidate physical systems \cite{NC2000, Qdot2009, Waveguides2011, Atomic-ensembles2013, Trapped-ions2013, NMR2013,SDL2010, Liu2013}, in particular, one can prepare and manipulate  multipartite entanglement in optical systems \cite{SZ1997, KLM2001, Kok2007, Pan2012, Li2014, LZAWN2013, RWD2015}.

Generally, a spontaneous parametric down-conversion (PDC) source \cite{PDC1995, SB2003} is capable of emitting pairs of strongly time-correlated photons in two spatial modes. As extensions of interest, with linear optics and nonlinear optical materials several schemes for creating multiphoton entangled states have been proposed \cite{GHZExperiment99, Pan2001, QE-LNumberP2004, WAZ2007, WLK2011, TMB2013, HDYG2015JPB}.
For a large number of  photons, however, there are some technological challenges such as probabilistic emission of PDC sources and imperfect detectors.
A feasible approach is to use the simple single-photon sources, instead of waiting the successive pairs, and quantum nondemolition (QND) measurement \cite{Barrett2005, MNBS2005, NM2004, MNS2005, Kok2008, DYG2013, DYG2014, HDYG2015OE} with weak Kerr nonlinearities.
Note that the Kerr nonlinearities \cite{Imoto1985, RV2005, Optical-Fiber-Kerr2009} are extremely weak and the order of magnitude of them is only $10^{-2}$ even by using electromagnetically induced transparency \cite{SI1996, LI2001}.
More recently, Shapiro \emph{et al} \cite{SR2007, DCS2014} showed that the causality-induced phase noise will preclude high-fidelity $\pi$-radian conditional phase shifts created by the cross-Kerr effect.
In these cases, with the increase of the number of photons it is usually more and more difficult to study multiphoton entanglement in the regime of weak nonlinearities.

In this paper, we focus on the exploration of multiphoton entangled states with linear optics and weak nonlinearities.
We show a quantum circuit to evolve multimode signal photons fed by a group of arbitrary single-photon states and  the coherent probe beam.
Particularly, there are only two specified but small phase shifts induced in the process of interaction with weak nonlinearities. This fruitful architecture allows us to explore multiphoton entangled states with a large number of photons but still in the regime of weak nonlinearities.

\section{Creation of multiphoton entangled states with linear optics and weak nonlinearities}

Before describing the proposed scheme, let us first give a brief introduction of the Kerr nonlinearities.
The nonlinear Kerr media can be used to induce a cross phase modulation with Hamiltonian of the form
$H =\hbar \chi a_s^\dag {a_s}a_p^\dag {a_p}$, where $\chi$ is the coupling constant and ${a_s}$ (${a_p}$) represents the annihilation operator for photons in the signal (probe) mode.
If we assume that the signal mode is initially described by the state
${\left| \psi \right\rangle _s} = {c_0} {\left| 0 \right\rangle _s} + {c_1}{\left| 1 \right\rangle _s}$ and the coherent probe beam is $\left| \alpha  \right\rangle _p$,
then after the Kerr interaction the whole system evolves as
${\text{e}^{\text{i} H t/\hbar }}{\left| \psi  \right\rangle _s}{\left| \alpha  \right\rangle _p} = c_0 {\left| 0 \right\rangle _s}{\left| \alpha  \right\rangle _p} + c_1 {\left| 1 \right\rangle _s}{\left|{\alpha {\text{e}^{\texttt{i}\theta }}} \right\rangle _p}$,
where $\theta  = \chi t$ with interaction time $t$.
In order to distinguish different cases, one may perform a homodyne measurement \cite{MNBS2005} on the probe beam with quadrature operator $\hat{x}(\phi)=a_p\text{e}^{\text{i}\phi}+a_p^\dag\text{e}^{-\text{i}\phi}$, where $\phi$ is a real constant. Especially for $\phi=0$, this operation is conventionally referred to as $X$ homodyne measurement; while for $\phi=\pi/2$, it is called $P$ homodyne measurement.

\begin{figure}
  \includegraphics[width=4.5in]{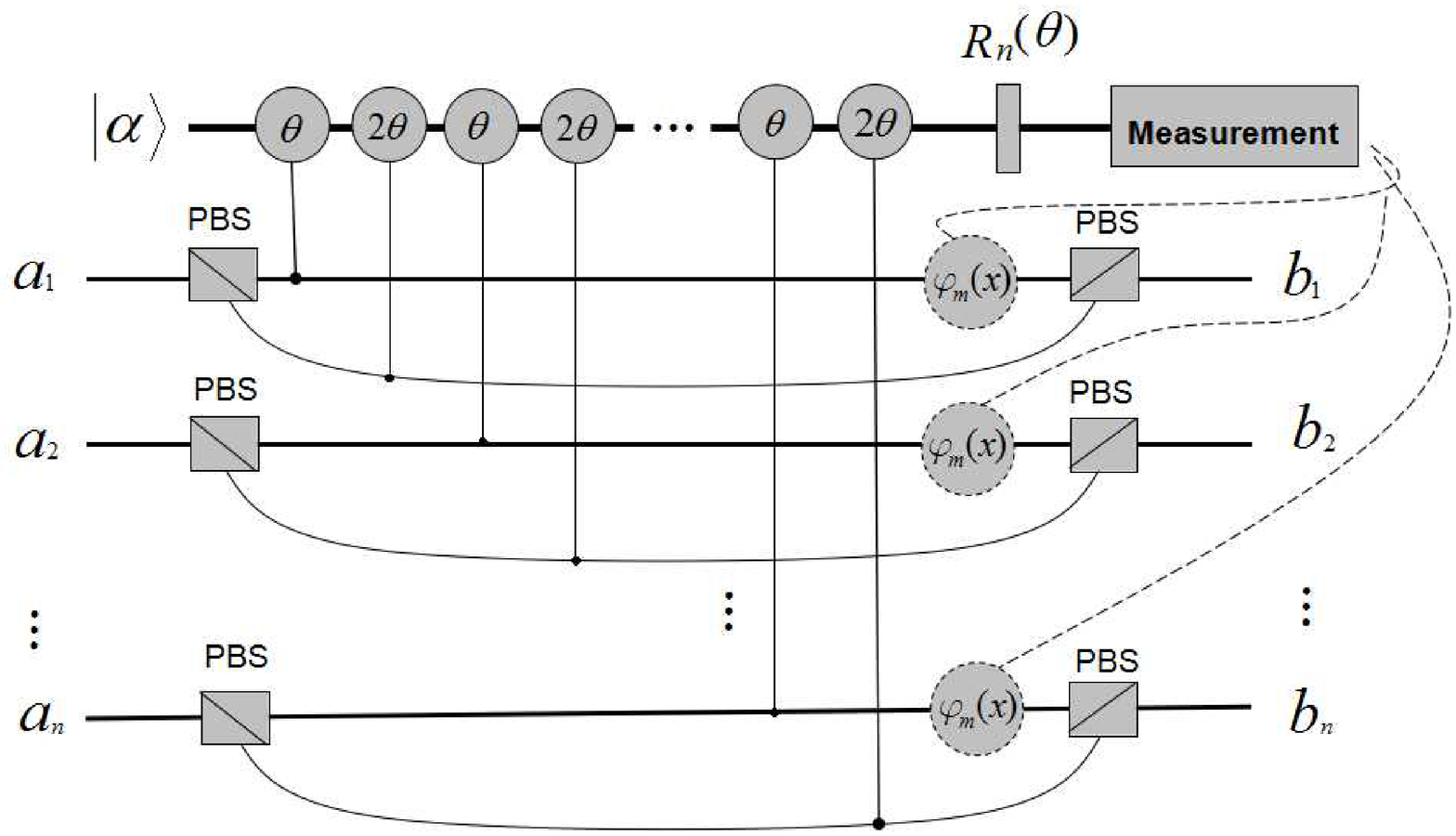}\\
  \caption{The schematic diagram of creating multiphoton entangled states with linear optics and weak nonlinearities. $a_{i}, i=1, 2, \cdots, n$ are input ports and each port is supplied with an arbitrary single-photon state, while $b_{i}, i=1,2,\cdots,n$ are the corresponding outputs, respectively.  $\theta$ and $2\theta$ represent phase shifts in the coherent probe beam $|\alpha\rangle$ induced by Kerr interaction between photons. $R_{n}(\theta)$ is a phase shift gate. Each ${\varphi _m}\left( x \right)$ represents a phase shift on the signal photons based on classical feed-forward information.}
  \label{multiphoton-entangler}
\end{figure}

Let $a_{i}, i=1, 2, \cdots, n$ represent input ports with respective spatial modes, namely signal modes, and $|\alpha\rangle$ is a coherent beam in probe mode. The setup of creating multiphoton entangled states is shown in Fig.\ref{multiphoton-entangler}.

Without loss of generality we may suppose that each input port is supplied with an arbitrary single-photon state.
Then, the total input state reads
\begin{eqnarray}\label{}
|\Psi_{\text{in}}\rangle|\alpha\rangle &=& (a_{00 \cdots 00}|H\rangle^{\otimes n}+\sum_{\text{Perm}}a_{100 \cdots 0}|H\rangle^{\otimes (n-1)} |V\rangle
+\sum_{\text{Perm}}a_{110 \cdots 0}|H\rangle^{\otimes (n-2)} |V\rangle^{\otimes 2} + \cdots \nonumber \\
& &
+\sum_{\text{Perm}}a_{1 \cdots 100}|H\rangle^{\otimes 2} |V\rangle^{\otimes (n-2)}
+ \sum_{\text{Perm}}a_{1 \cdots 110}|H\rangle |V\rangle^{\otimes (n-1)}
+ a_{11 \cdots 11}|V\rangle^{\otimes n})|\alpha\rangle,
\end{eqnarray}
where $a_{i_1i_2 \cdots i_n}$, $i_1,i_2, \cdots, i_n\in \{0,1\}$, are complex coefficients satisfying the normalization condition $\sum_{i_1,i_2, \cdots, i_n=0}^1|a_{i_1i_2 \cdots i_n}|^{2}=1$
and $\sum_{\text{Perm}}$ denotes the sum over all possible permutations of the signal modes, for example, $\sum_{\text{Perm}}a_{100 \cdots 0}|H\rangle^{\otimes (n-1)} |V\rangle \equiv a_{100 \cdots 0}|V\rangle_{a_1}|H\rangle^{\otimes (n-1)}_{a_2, \cdots, a_n} + a_{010 \cdots 0}|H\rangle_{a_1} |V\rangle_{a_2} |H\rangle^{\otimes (n-2)}_{a_3, \cdots, a_n}+ \cdots + a_{0 \cdots 010}|H\rangle^{\otimes (n-2)}_{a_1, \cdots, a_{n-2}}|V\rangle_{a_{n-1}}|H\rangle_{a_n} + a_{0 \cdots 001}|H\rangle^{\otimes (n-1)}_{a_1, \cdots, a_{n-1}}|V\rangle_{a_n}$.
Each polarizing beam splitter (PBS) is used to transmit $|H\rangle$ polarization photons and reflect $|V\rangle$ polarization photons.
When the signal photons travel to the PBSs, they will be individually split into two spatial modes and then interact with the nonlinear media so that pairs of phase shifts $\theta$ and $2\theta$ are induced on the coherent probe beam, respectively.
We here introduce a single phase gate $R_{n}(\theta)=-3n\theta/2$ so as to implement the next $X$ homodyne measurement on the probe beam.
${\varphi _m}\left( x \right)$, $m=1, 2 ,\cdots, n/2$ for even $n$ and $m=1/2, 3/2 ,\cdots, n/2$ for odd $n$, are phase shifts on the signal photons based on the measured values of $x$ via the classical feed-forward information.
At last, at the ports $b_{i}, i=1,2,\cdots,n$ one may obtain $n/2+1$ output states for even $n$ or $(n+1)/2$ states for odd $n$.

Now we describe our method in details. For $n$ is even, after the interaction between the photons with Kerr media and followed by the action of the phase gate, the combined system evolves as
\begin{eqnarray}\label{}
|\Psi_{\text{ck}}\rangle &=& a_{00 \cdots 00}|H\rangle^{\otimes n} |\alpha \text{e}^{-(n/2)\text{i}\theta}\rangle
+(\sum_{\text{Perm}}a_{100 \cdots 0}|H\rangle^{\otimes (n-1)} |V\rangle) |\alpha \text{e}^{-(n/2-1)\text{i}\theta}\rangle  \nonumber \\
& & +(\sum_{\text{Perm}}a_{110 \cdots 0}|H\rangle^{\otimes (n-2)} |V\rangle^{\otimes 2}) |\alpha \text{e}^{-(n/2-2)\text{i}\theta}\rangle + \cdots +
(\sum_{\text{Perm}}a_{\underbrace{11\cdots1}_{n/2}\underbrace{00\cdots0}_{n/2}}|H\rangle^{\otimes n/2} |V\rangle^{\otimes n/2} )|\alpha \rangle
\nonumber \\
& & + \cdots +(\sum_{\text{Perm}}a_{1 \cdots 100}|H\rangle^{\otimes 2} |V\rangle^{\otimes (n-2)} )|\alpha \text{e}^{(n/2-2)\text{i}\theta}\rangle  +(\sum_{\text{Perm}}a_{11 \cdots 10}|H\rangle |V\rangle^{\otimes (n-1)} )|\alpha \text{e}^{(n/2-1)\text{i}\theta}\rangle \nonumber \\
& &+ a_{11 \cdots 11}|V\rangle^{\otimes n} |\alpha \text{e}^{(n/2)\text{i}\theta}\rangle.
\end{eqnarray}
In order to create the desired multiphoton entangled states, we here perform an $X$ homodyne measurement \cite{Barrett2005, HDYG2015OE} on the probe beam. If the value $x$ of the $X$ homodyne measurement is obtained, then the signal photons become
\begin{eqnarray}\label{}
|\Psi_{x}\rangle &=&
f({x,\alpha}) \times \sum_{\text{Perm}}a_{\underbrace{11\cdots1}_{n/2}\underbrace{00\cdots0}_{n/2}}|H\rangle^{\otimes n/2} |V\rangle^{\otimes n/2} \nonumber \\
& & + f({x,\alpha \cos \theta }) \times (\text{e}^{-\text{i}\phi _{1}(x)}\sum_{\text{Perm}}a_{\underbrace{11\cdots1}_{n/2-1}\underbrace{00\cdots0}_{n/2+1}}|H\rangle^{\otimes (n/2+1)} |V\rangle^{\otimes (n/2-1)} \nonumber \\
& &+ \text{e}^{\text{i}\phi _{1}(x)}\sum_{\text{Perm}}a_{\underbrace{11\cdots1}_{n/2+1}\underbrace{00\cdots0}_{n/2-1}}|H\rangle^{\otimes (n/2-1)} |V\rangle^{\otimes (n/2+1)})+ \cdots \nonumber \\
& & + f({x,\alpha \cos [(n/2-1)\theta] }) \times ( \text{e}^{-\text{i}\phi _{n/2-1}(x)}\sum_{\text{Perm}}a_{100 \cdots 0} |H\rangle^{\otimes (n-1)} |V\rangle
 + \text{e}^{\text{i}\phi _{n/2-1}(x)}\sum_{\text{Perm}}a_{11 \cdots 10}  |H\rangle |V\rangle^{\otimes (n-1)})\nonumber \\
& & + f({x,\alpha \cos (n\theta/2) }) \times ( \text{e}^{-\text{i}\phi _{n/2}(x)} a_{00 \cdots 00} |H\rangle^{\otimes n} + \text{e}^{\text{i}\phi _{n/2}(x)} a_{11 \cdots 11}  |V\rangle^{\otimes n}),
\end{eqnarray}
where $f(x, \alpha \cos (m \theta)) = {\left( {2\pi } \right)^{ - 1/4}}{\text{e}^{ - [{x - 2{\alpha \cos (m \theta)}}]^2/4}}$, $m=0, 1, 2, \cdots, n/2$, are respectively Gaussian curves which are associated with the probability amplitudes of the outputs, and ${\phi _m}\left( x \right) = \alpha \sin (m\theta)[x - 2\alpha \cos (m \theta)]  \bmod 2\pi$, $m=1, 2, \cdots, n/2$, are respectively phase factors based on the values of the $X$ homodyne measurement.
Note that the peaks of these Gaussian distributions locate at $2\alpha \cos (m\theta)$. Thus, the midpoints between two neighboring peaks ${x_{{m_{k}}}} =2\alpha \cos (\theta/2) \cos [(k-1/2)\theta]$ and the distances of two nearby peaks ${x_{{d_k}}} = 2\alpha \{{\cos [(k-1)\theta]  - \cos(k\theta) } \}$ with $k=1, 2, \cdots, n/2$. Obviously, with these $n/2$ midpoint values ${x_{{m_{k}}}}$ there exist $n/2+1$ intervals and each interval corresponds to an output state.

We now consider the phase shifts $\varphi_m(x)$. The signal photon  evolves as $\hat{b}_{i}^\dag=\text{e}^{\text{i}\varphi _{m}(x)}\hat{a}_{i}^\dag$, $i=1, 2, \cdots, n$. A straightforward calculation shows that $\varphi _{m}(x) = \alpha \sin (m\theta) [ {x - 2\alpha \cos (m\theta) } ]/m \bmod 2\pi$.
After  these feed-forward phase shifts have been implemented and the signal photons pass through the PBSs,  one can obtain the desired states as follows.
Clearly, for $x < {x_{{m_{n/2}}}}$ we have $a_{00 \cdots 00}|H\rangle^{\otimes n} + a_{11 \cdots 11} |V\rangle^{\otimes n}$; for $x_{{m_{k+1}}} < x < x_{{m_{k}}}, k=1, 2, \cdots, n/2-1$, we obtain the states $\sum_{\text{Perm}}a_{\underbrace{11\cdots1}_{n/2-k}\underbrace{00\cdots0}_{n/2+k}}|H\rangle^{\otimes (n/2+k)} |V\rangle^{\otimes (n/2-k)} + \sum_{\text{Perm}}a_{\underbrace{11\cdots1}_{n/2+k}\underbrace{00\cdots0}_{n/2-k}}|H\rangle^{\otimes (n/2-k)} |V\rangle^{\otimes (n/2+k)}$; and for $x > {x_{{m_{1}}}}$ we obtain the state $\sum_{\text{Perm}}a_{\underbrace{11\cdots1}_{n/2}\underbrace{00\cdots0}_{n/2}}|H\rangle^{\otimes n/2} |V\rangle^{\otimes n/2}$.

Similarly, for odd $n$, we have
\begin{eqnarray}\label{}
|\Psi_{\text{ck}}\rangle &=& a_{00 \cdots 00}|H\rangle^{\otimes n} |\alpha \text{e}^{-(n/2)\text{i}\theta}\rangle
+ (\sum_{\text{Perm}}a_{100 \cdots 0}|H\rangle^{\otimes (n-1)} |V\rangle )|\alpha \text{e}^{-(n/2-1)\text{i}\theta}\rangle  \nonumber \\
& & +(\sum_{\text{Perm}}a_{110 \cdots 0}|H\rangle^{\otimes (n-2)} |V\rangle^{\otimes2} )|\alpha \text{e}^{-(n/2-2)\text{i}\theta}\rangle + \cdots
+(\sum_{\text{Perm}}a_{\underbrace{11\cdots1}_{(n-1)/2}\underbrace{00\cdots0}_{(n+1)/2}}|H\rangle^{\otimes (n+1)/2} |V\rangle^{\otimes (n-1)/2} )|\alpha \text{e}^{-1/2\text{i}\theta} \rangle \nonumber \\
& &+(\sum_{\text{Perm}}a_{\underbrace{11\cdots1}_{(n+1)/2}\underbrace{00\cdots0}_{(n-1)/2}}|H\rangle^{\otimes (n-1)/2} |V\rangle^{\otimes (n+1)/2}) |\alpha \text{e}^{1/2\text{i}\theta} \rangle
+ \cdots +(\sum_{\text{Perm}}a_{11 \cdots 100}|H\rangle^{\otimes2} |V\rangle^{\otimes (n-2)}) |\alpha \text{e}^{(n/2-2)\text{i}\theta}\rangle \nonumber \\
& &  +(\sum_{\text{Perm}}a_{11 \cdots 10}|H\rangle |V\rangle^{\otimes n-1} )|\alpha \text{e}^{(n/2-1)\text{i}\theta}\rangle
+ a_{11 \cdots 11}|V\rangle^{\otimes n} |\alpha \text{e}^{(n/2)\text{i}\theta}\rangle.
\end{eqnarray}
Also,
\begin{eqnarray}\label{}
|\Psi_{x}\rangle &=& f(x,\alpha \cos (\theta/2))  \nonumber \\
& & \times (\text{e}^{-\text{i}\phi _{1/2}(x)}\sum_{\text{Perm}}a_{\underbrace{11\cdots1}_{(n-1)/2}\underbrace{00\cdots0}_{(n+1)/2}}  |H\rangle^{\otimes (n+1)/2}|V\rangle^{\otimes (n-1)/2}+ \text{e}^{\text{i}\phi _{1/2}(x)} \sum_{\text{Perm}}a_{\underbrace{11\cdots1}_{(n+1)/2}\underbrace{00\cdots0}_{(n-1)/2}} |H\rangle^{\otimes (n-1)/2} |V\rangle^{\otimes (n+1)/2}) \nonumber \\
& & + f(x,\alpha \cos (3\theta/2)) \nonumber \\
& &\times (\text{e}^{-\text{i}\phi _{3/2}(x)}\sum_{\text{Perm}}a_{\underbrace{11\cdots1}_{(n-3)/2}\underbrace{00\cdots0}_{(n+3)/2}}|H\rangle^{\otimes (n+3)/2} |V\rangle^{\otimes (n-3)/2} + \text{e}^{\text{i}\phi _{3/2}(x)}\sum_{\text{Perm}}a_{\underbrace{11\cdots1}_{(n+3)/2}\underbrace{00\cdots0}_{(n-3)/2}}|H\rangle^{\otimes (n-3)/2} |V\rangle^{\otimes (n+3)/2})\nonumber \\
& & + \cdots + f({x,\alpha \cos [(n/2-1)\theta] }) \times ( \text{e}^{-\text{i}\phi _{n/2-1}(x)}\sum_{\text{Perm}}a_{100 \cdots 0} |H\rangle^{\otimes (n-1)} |V\rangle + \text{e}^{\text{i}\phi _{n/2-1}(x)}\sum_{\text{Perm}}a_{11 \cdots 10}  |H\rangle |V\rangle^{\otimes (n-1)})\nonumber \\
& & + f({x,\alpha \cos (n\theta/2) }) \times (\text{e}^{-\text{i}\phi _{n/2}(x)} a_{00 \cdots 00}  |H\rangle^{\otimes n} + \text{e}^{\text{i}\phi _{n/2}(x)} a_{11 \cdots 11}  |V\rangle^{\otimes n}).
\end{eqnarray}
Here, the functions $f(x,\alpha \cos (m\theta))$, phase shifts $\phi_{m}(x)$ and $\varphi_{m}(x)$ are approximately the same as those described for even $n$, except for $m=1/2, 3/2, \cdots, {n/2}$, and then the similar results hold for the midpoints ${x_{{m_{k}}}}$ and the distances ${x_{{d_k}}}$ with $k=3/2, 5/2, \cdots, n/2$.
Of course, with $(n-1)/2$ midpoint values ${x_{{m_{k}}}}$ there may be $(n+1)/2$ output states;
that is, one can obtain the states $a_{00 \cdots 00}|H\rangle^{\otimes n} + a_{11 \cdots 11} |V\rangle^{\otimes n}$ for $x < {x_{{m_{n/2}}}}$,
$ \sum_{\text{Perm}}a_{\underbrace{11\cdots1}_{n/2-k}\underbrace{00\cdots0}_{n/2+k}}|H\rangle^{\otimes (n/2+k)} |V\rangle^{\otimes (n/2-k)}
+ \sum_{\text{Perm}}a_{\underbrace{11\cdots1}_{n/2+k}\underbrace{00\cdots0}_{n/2-k}}|H\rangle^{\otimes (n/2-k)} |V\rangle^{\otimes (n/2+k)}$
for $x_{{m_{k+1}}} < x < x_{{m_{k}}}, k=3/2, \cdots, n/2-1$,
and $\sum_{\text{Perm}}a_{\underbrace{11\cdots1}_{(n-1)/2}\underbrace{00\cdots0}_{(n+1)/2}}|H\rangle^{\otimes (n+1)/2} |V\rangle^{\otimes (n-1)/2}+ \sum_{\text{Perm}}a_{\underbrace{11\cdots1}_{(n+1)/2}\underbrace{00\cdots0}_{(n-1)/2}}|H\rangle^{\otimes (n-1)/2} |V\rangle^{\otimes (n+1)/2}$ for $x > {x_{{m_{3/2}}}}$.

As an example of the applications of interest for the present scheme, we  introduce a class of remarkable multipartite entangled states
\begin{equation}\label{}
|\Psi_n^k\rangle=\frac{1}{\sqrt{2}}(|\psi_n^k\rangle+|\widetilde{\psi}_n^k\rangle),
\end{equation}
where $|\psi_n^k\rangle= \frac{1}{\sqrt{\binom{n}{n/2-k}}} \sum_{\text{Perm}}|0\rangle^{\otimes {(n/2+k)}} |1\rangle^{\otimes {(n/2-k)}}$ and $|\widetilde{\psi}_n^k\rangle=\frac{1}{\sqrt{\binom{n}{n/2-k}}} \sum_{\text{Perm}}|0\rangle^{\otimes {(n/2-k)}} |1\rangle^{\otimes {(n/2+k)}}$ are two orthonormal states, namely Dicke states \cite{Dicke1954, TZBSA2007}. In view of its ``catness'', the state $|\Psi_n^k\rangle$ can be referred to as \emph{cat-like state},
and especially for $k=n/2$ it can be expressed as the canonical $n$-partite Greenberger-Horne-Zeilinger (GHZ) state.
In the present scheme, obviously, for $a_{i_1i_2 \cdots i_n}=1/\sqrt{2^n}$, $i_1,i_2, \cdots, i_n=0,1$, we can obtain these cat-like states
with $k=0,1,\cdots, n/2$ for even $n$ and $k=1/2,3/2,\cdots, n/2$ for odd $n$, where the qubits are encoded with the polarization modes $ \left|H\right\rangle \equiv \left|0\right\rangle $ and $\left|V \right\rangle \equiv \left|1 \right\rangle$. Of course, more generally, we may project out a group of multiphoton entangled states involving generalized Dicke states.

\section{discussion and summary}

There are two models commonly employed in the process of Kerr interaction, single-mode model and continuous-time multi-mode model \cite{SR2007}. The former implies that one may ignore the temporal behavior of the optical pulses but the latter is causal, non-instantaneous model involving phase noise.
It has been shown that \cite{DCS2014} this causality-induced phase noise will preclude the possibility of high-fidelity CPHASE gates created by the cross-Kerr effect in optical fiber. To solve this problem, one may need to find an optimum response function for the available medium, or to exploit more favorable systems, such as cavitylike systems \cite{CCS2013}. After all, the ultimate possible performance of Kerr interaction with a larger system is an interesting open issue.
In the present scheme, we restrict ourselves to ignoring the phase noise and concentrate mainly on showing a method for exploring multiphoton entangled states in the regime of weak cross-Kerr nonlinearities, i.e. $\theta \ll \pi$.

It is worth noting that, there are only small phase shifts $\theta$ and $2\theta$ instead of a series of related functions of the number of photons  in the process of interaction with Kerr nonlinearities. This implies that the present scheme is quite tractable especially for creating entangled states with a larger number of photons.
In addition,  the error probabilities $\varepsilon_k$  are  $\textrm{erfc}\left( {{x_{d_k}}/2\sqrt 2 } \right)/2$, which come from small overlaps between two neighboring curves. Considering the distances of two nearby peaks ${x_{{d_k}}} \approx (2k-1)\alpha {\theta ^2}$ with $k=1, 2, \cdots, n/2$ for even $n$ and $k=3/2, 5/2, \cdots, n/2$ for odd $n$, the maximal error probability $\varepsilon_{\text{max}} = \textrm{erfc}\left( {{\alpha {\theta ^2}}/2\sqrt 2 } \right)/2$, which is exactly the result described by Nemoto and Munro in \cite{NM2004}.
Obviously, the error probabilities in our scheme are no more than that one even when the number of photons  is large.

Evidently, three aspects are noteworthy in the present framework. First, since there are no large phase shifts in the interacting process with weak Kerr nonlinearities, our scheme is more feasible compared with the previous schemes.
Second, the system is measured only once with a small error probability and it means that the present scheme might be realized near deterministically.
Finally, the fruitful architecture allows us to explore a group of multiphoton entangled states involving a large number of photons,  i.e., to produce entangled states approaching the macroscopic domain.

\section{Acknowledgements}
This work was supported by the National Natural Science Foundation of China under Grant Nos: 11475054, 11371005, Hebei Natural Science Foundation of China under Grant No:  A2014205060, the Research Project of Science and Technology in Higher Education of Hebei Province of China under Grant No: Z2015188, Langfang Key Technology Research and Development Program of China under Grant No: 2014011002.


\begin{thebibliography}{99}

\bibitem{HHHH2009} R. Horodecki, P. Horodecki, M. Horodecki, K. Horodecki, Rev. Mod. Phys. {\bf 81}, 865 (2009).
\bibitem{GT2009} O. G$\ddot{\textrm{u}}$hne and G. T$\acute{\textrm{o}}$th, Phys. Rep. {\bf 474}, 1 (2009).
\bibitem{YGC2011} F. L. Yan, T. Gao, and E. Chitambar, Phys. Rev. A {\bf 83}, 022319 (2011).
\bibitem{GYE2014} T. Gao, F. L. Yan, and S. J. van Enk, Phys. Rev. Lett. {\bf 112}, 180501 (2014).


\bibitem{NC2000} M. A. Nielsen and I. L. Chuang, \textsl{Quantum Computation and Quantum Information} (Cambridge University Press, Cambridge, 2000).
\bibitem{Qdot2009} E. Waks and C. Monroe, Phys. Rev. A {\bf 80}, 062330 (2009).
\bibitem{Waveguides2011} A. Gonzalez-Tudela, D. Martin-Cano, E. Moreno, L. Martin-Moreno, C. Tejedor, and F. J. Garcia-Vidal, Phys. Rev. Lett. {\bf 106}, 020501 (2011).
\bibitem{Atomic-ensembles2013} D. Shwa, R. D. Cohen, A. Retzker, and N. Katz, Phys. Rev. A {\bf 88}, 063844 (2013).
\bibitem{Trapped-ions2013} B. Casabone, A. Stute, K. Friebe, B. Brandst$\ddot{\textrm{a}}$tter, K. Sch$\ddot{\textrm{u}}$ppert, R. Blatt, and T. E. Northup, Phys. Rev. Lett. {\bf 111}, 100505 (2013).
\bibitem{NMR2013} G. R. Feng, G. L. Long, and R. Laflamme, Phys. Rev. A {\bf 88}, 022305 (2013).
\bibitem{SDL2010} Y. B. Sheng, F. G. Deng, and  G. L. Long, Phys. Rev. A {\bf 82}, 032318 (2010).
\bibitem{Liu2013} S. P. Liu, J. H. Li, R. Yu, and  Y.  Wu, Phys. Rev. A {\bf 87}, 062316 (2013).

\bibitem{SZ1997} M. O. Scully and M. S. Zubairy, \textsl{Quantum Optics} (Cambridge University Press, Cambridge, 1997).
\bibitem{KLM2001} E. Knill, R. Laflamme, and G. J. Milburn, Nature (London)  {\bf 409}, 46 (2001).
\bibitem{Kok2007} P. Kok, W. J. Munro, K. Nemoto, T. C. Ralph, J. P. Dowling, and G. J. Milburn, Rev. Mod. Phys. {\bf 79}, 135 (2007).
\bibitem{Pan2012} J. W. Pan, Z. B. Chen, C. Y. Lu, H. Weinfurter, A. Zeilinger, and M. $\dot{\textrm{Z}}$kowski, Rev. Mod. Phys.  {\bf 84}, 777 (2012).

\bibitem{Li2014} J. H. Li, R. Yu, C. L. Ding, W. Wang, and Y. Wu,  Phys. Rev. A {\bf 90}, 033830 (2014).
\bibitem{LZAWN2013} X. Y. L$\ddot{\textrm{u}}$, W. M. Zhang, S. Ashhab, Y. Wu, and F. Nori, Sci. Rep. {\bf 3}, 2943 (2013).
\bibitem{RWD2015} B. C. Ren, G. Y. Wang, and F. G. Deng, Phys. Rev. A {\bf 91}, 032328 (2015).

\bibitem{PDC1995} P. G. Kwiat, K. Mattle, H. Weinfurter, A. Zeilinger, A. V. Sergienko, and Y. Shih, Phys. Rev. Lett. {\bf 75}, 4337 (1995).
\bibitem{SB2003} C. Simon and D. Bouwmeester, Phys. Rev. Lett. {\bf 91}, 053601 (2003).

\bibitem{GHZExperiment99} D. Bouwmeester, J. W. Pan, M. Daniell, H. Weinfurter, and A. Zeilinger,  Phys. Rev. Lett. {\bf 82}, 1345 (1999).
\bibitem{Pan2001} J. W. Pan, M. Daniell, S. Gasparoni, G. Weihs, and A. Zeilinger, Phys. Rev. Lett. {\bf 86}, 4435 (2001).
\bibitem{QE-LNumberP2004} H. S. Eisenberg, G. Khoury, G. A. Durkin, C. Simon, and D. Bouwmeester, Phys. Rev. Lett. {\bf 93}, 193901 (2004).
\bibitem{WAZ2007} P. Walther, M. Aspelmeyer, and A. Zeilinger, Phys. Rev. A {\bf 75}, 012313 (2007).
\bibitem{WLK2011} T. C. Wei, J. Lavoie, and R. Kaltenbaek, Phys. Rev. A {\bf 83}, 033839 (2011).
\bibitem{TMB2013} M. C. Tichy, F. Mintert, and A. Buchleitner, Phys. Rev. A {\bf 87}, 022319 (2013).
\bibitem{HDYG2015JPB} Y. Q. He, D. Ding, F. L. Yan, and T. Gao, J. Phys. B: At. Mol. Opt. Phys. {\bf 48}, 055501 (2015).

\bibitem{Barrett2005} S. D. Barrett, P. Kok, K. Nemoto, R. G. Beausoleil, W. J. Munro, and T. P. Spiller, Phys. Rev. A {\bf 71}, 060302 (2005).
\bibitem{MNBS2005} W. J. Munro, K. Nemoto, R. G. Beausoleil, and T. P. Spiller,  Phys. Rev. A {\bf 71}, 033819 (2005).
\bibitem{NM2004} K. Nemoto and W. J. Munro, Phys. Rev. Lett. {\bf 93}, 250502 (2004).
\bibitem{MNS2005} W. J. Munro, K. Nemoto, and T. P. Spiller, New J. Phys. {\bf 7}, 137 (2005).
\bibitem{Kok2008} P. Kok, Phys. Rev. A {\bf 77}, 013808 (2008).
\bibitem{DYG2013} D. Ding, F. L. Yan, and T. Gao,  J. Opt. Soc. Am. B {\bf 30}, 3075 (2013).
\bibitem{DYG2014} D. Ding, F. L. Yan, and T. Gao, Sci. China-Phys. Mech. Astron. {\bf 57}, 2098 (2014).
\bibitem{HDYG2015OE} Y. Q. He, D. Ding, F. L. Yan, and T. Gao, Opt. Express {\bf 23}, 21671 (2015).

\bibitem{Imoto1985} N. Imoto, H. A. Haus, and Y. Yamamoto, Phys. Rev. A {\bf 32}, 2287 (1985).

\bibitem{RV2005} H. Rokhsari and K. J. Vahala, Opt. Lett. {\bf 30}, 427 (2005).
\bibitem{Optical-Fiber-Kerr2009} N. Matsuda, R. Shimizu, Y. Mitsumori, H. Kosaka, and K. Edamatsu, Nature Photon {\bf 3}, 95 (2009).

\bibitem{SI1996} H. Schmidt and A. Imamo$\check{\textrm{g}}$lu, Opt. Lett. {\bf 21}, 1936 (1996).
\bibitem{LI2001} D. Lukin and A. Imamo$\check{\textrm{g}}$lu, Nature (London) {\bf 413}, 273 (2001).

\bibitem{SR2007} J. H. Shapiro and M. Razavi, New J. Phys. {\bf 9}, 16 (2007).
\bibitem{DCS2014} J. Dove, C. Chudzicki, and J. H. Shapiro,  Phys. Rev. A {\bf 90}, 062314 (2014).

\bibitem{Dicke1954} R. H. Dicke, Phys. Rev. {\bf 93}, 99 (1954).
\bibitem{TZBSA2007} C. Thiel, J. von Zanthier, T. Bastin, E. Solano, and G. S. Agarwal, Phys. Rev. Lett. {\bf 99}, 193602 (2007).

\bibitem{CCS2013} C. Chudzicki, I. L. Chuang, and J. H. Shapiro, Phys. Rev. A {\bf 87}, 042325 (2013).


\end{thebibliography}
\end{document}